\documentclass[amsmath,amssymb,aip,apl,reprint]{revtex4-2}

\usepackage{graphicx}
\usepackage{dcolumn}
\usepackage{bm}
\usepackage[utf8]{inputenc}
\usepackage[T1]{fontenc}
\usepackage{mathptmx}
\usepackage{etoolbox}
\usepackage{mathtools}

\begin{document}

\title{Free expansion of a charged nanoparticle via electrostatic compensation}
\author{David Steiner}
\affiliation{Faculty of Physics, Vienna Center for Quantum Science and Technology (VCQ), University of Vienna, Boltzmanngasse 5, A-1090 Vienna, Austria}
\affiliation{Institute for Quantum Optics and Quantum Information (IQOQI) Vienna, Austrian Academy of Sciences, A-1090 Vienna, Austria}
\author{Yaakov Y. Fein}
\email{yaakov.fein@univie.ac.at}
\affiliation{Faculty of Physics, Vienna Center for Quantum Science and Technology (VCQ), University of Vienna, Boltzmanngasse 5, A-1090 Vienna, Austria}
\author{Gregor Meier} 
\affiliation{Faculty of Physics, Vienna Center for Quantum Science and Technology (VCQ), University of Vienna, Boltzmanngasse 5, A-1090 Vienna, Austria}
\author{Stefan Lindner} 
\affiliation{Faculty of Physics, Vienna Center for Quantum Science and Technology (VCQ), University of Vienna, Boltzmanngasse 5, A-1090 Vienna, Austria}
\author{Paul Juschitz} 
\affiliation{Faculty of Physics, Vienna Center for Quantum Science and Technology (VCQ), University of Vienna, Boltzmanngasse 5, A-1090 Vienna, Austria}
\author{Mario A. Ciampini} 
\affiliation{Faculty of Physics, Vienna Center for Quantum Science and Technology (VCQ), University of Vienna, Boltzmanngasse 5, A-1090 Vienna, Austria}
\author{Markus Aspelmeyer}
\affiliation{Faculty of Physics, Vienna Center for Quantum Science and Technology (VCQ), University of Vienna, Boltzmanngasse 5, A-1090 Vienna, Austria}
\affiliation{Institute for Quantum Optics and Quantum Information (IQOQI) Vienna, Austrian Academy of Sciences, A-1090 Vienna, Austria}
\author{Nikolai Kiesel}
\affiliation{Faculty of Physics, Vienna Center for Quantum Science and Technology (VCQ), University of Vienna, Boltzmanngasse 5, A-1090 Vienna, Austria}

\date{\today}

\begin{abstract}
Coherent wavepacket expansion is a key component of recent proposals aiming to create non-classical states of a levitated dielectric nanoparticle.
Free evolution, i.e., releasing the particle from its harmonic trapping potential and allowing its position variance to grow, is a simple but effective expansion scheme, but requires accurate force compensation to avoid significant mean displacements of the wavepacket during the free evolution.
Here, using an optical trap-release-recapture sequence, we demonstrate an electrostatic compensation technique that enables free evolution without significant mean displacement in 3D, effectively compensating both gravity and stray electric fields. 
To achieve 100 microsecond free evolution times with charged particles, we developed methods to map and correct for force cross-talk, as well as to control the environmental charge state. 
Combined with a low decoherence environment, our approach enables the preparation of largely delocalized states without the need for long freefall trajectories or a low-gravity environment.
\end{abstract}

\maketitle

%%% INTRO %%%
Optically trapped dielectric nanoparticles are a promising platform for the exploration of macroscopic quantum physics~\cite{Millen2020,levitoreview}. 
A general challenge for manipulating such macroscopic oscillators is their small ground-state spatial extent, as given by the zero-point motion, $z_{\text{zp}} = \sqrt{\hbar/2m\Omega_z}$, with $\Omega_z$ the mechanical frequency along the optical axis ($z$) and $m$ the particle mass.
For typical optically trapped nanoparticles with diameters of 100\,nm, the zero-point motion of the axial mode is of order 10\,pm.
State expansion is therefore an integral step in recent proposals to demonstrate nanoparticle center-of-mass interference~\cite{Bateman2014,Neumeier2024,Roda_Llordes2024} in order for the particle wavepacket to sufficiently interact with a non-harmonic potential landscape.
State expansion can also be implemented for the purpose of time-of-flight tomography, as demonstrated with atoms and nanoparticles~\cite{Brown2023,kamba2025}, and as suggested in interferometry protocols~\cite{Bateman2014,Neumeier2024,Roda_Llordes2024}.
Coherent state expansion from a low starting occupation can also lead to squeezing of the mechanical mode~\cite{rossi2024,kamba2025}, which is both of fundamental interest and a practical tool in searches for new physics~\cite{Carney2021}.

Free expansion, i.e. evolution in the absence of a harmonic trapping potential, is a straightforward method for wavepacket expansion.
Releasing an non-squeezed thermal state with initial mean occupancy $\bar{n}_z$ for a time $\tau$ will coherently increase the axial position variance according to 
\begin{equation}
\label{eq:free}
    \sigma_z^2(\tau) =  z_{\text{zp}}^2(2\bar{n}_z+1)(1+\Omega_z^2 \tau^2).
\end{equation}
From here on, quantities ($\Omega$, $n$, etc.) without a subscript will refer to the axial mode unless otherwise specified. 

%%% state of the art %%% 
The duration of the free evolution $\tau$, and thereby the expansion, will eventually be limited by decoherence, with the strongest restriction typically coming from collisions with residual gas molecules. 
While vacuum levels below $10^{-10}$\,mbar can enable collision-free times of up to milliseconds, long free evolutions pose a number of experimental challenges even before this timescale.

Unlike in a harmonic trap, a free particle will be significantly displaced due to even small forces such as gravity.
Large displacements with respect to the trap center are detrimental because, in most cases, the free evolution will be followed by recapture and measurement in a harmonic potential. 
Even if the particle is successfully recaptured after the free evolution, a large increased motional amplitude resulting from a mean displacement from the trap center will lead to probing of trap nonlinearities, motional cross-coupling, and detector saturation. 
The use of charged particles, as required for some feedback-cooling techniques and potential landscape control, further complicates matters, since stray electric fields will displace the particle in 3D~\cite{Hebestreit2018}. 
The longest free evolutions reported with optically trapped nanoparticles have been of order 100\,µs~\cite{Hebestreit2018,Kamba2023,mattana2025}, but this has been restricted to neutral or nearly neutral particles.

%%% Alternative approaches %%%
As an alternative to free expansion, one can employ inverted harmonic potentials to accelerate the wavepacket expansion~\cite{Romero-Isart2017,Pino2018,Duchan2025,tomassi2025}, although there are experimental challenges both from the control perspective~\cite{Dago24} and in maintaining coherence~\cite{Weiss2021,tomassi2025}. 
Another option is to employ "frequency jumping", in which the nanoparticle is transferred between a stiff and a weak trap~\cite{Bonvin2024,rossi2024,Duchan2025,kamba2025}, which additionally offers a "safety net" to avoid particle loss and a detection during the evolution in the weaker trap.
Finally, there have been suggestions to perform the free expansion in a free-falling frame, e.g. using molecular fountains~\cite{Kialka2022}, drop towers~\cite{prakash2025} or even a dedicated satellite~\cite{Gasbarri2021}. 
However, free-falling schemes will likely still have to cope with non-gravitational residual forces, while frequency-jump protocols suffer from reduced state expansion and an increased sensitivity to noise~\cite{Bonvin2024} compared to free expansion. 

Here, we implement electrostatic force compensation to levitate a charged particle during free evolutions of order 100\,µs.
We measure the background forces experienced by the particle in situ and then apply compensating forces via 3D electrodes during the free expansion.
Unlike schemes that allow the particle to fall in gravity~\cite{Hebestreit2018,Kamba2023,mattana2025}, we keep the mean position of the particle in the lab frame constant during the free expansion, at the cost of also having to measure and compensate for background electric fields. 
The technique works with a range of particle charge states (typically, but not limited to, tens of elementary charges) and compensates both electrostatic and non-electrostatic forces.

\begin{figure}
    \centering
    \includegraphics[width=1\linewidth]{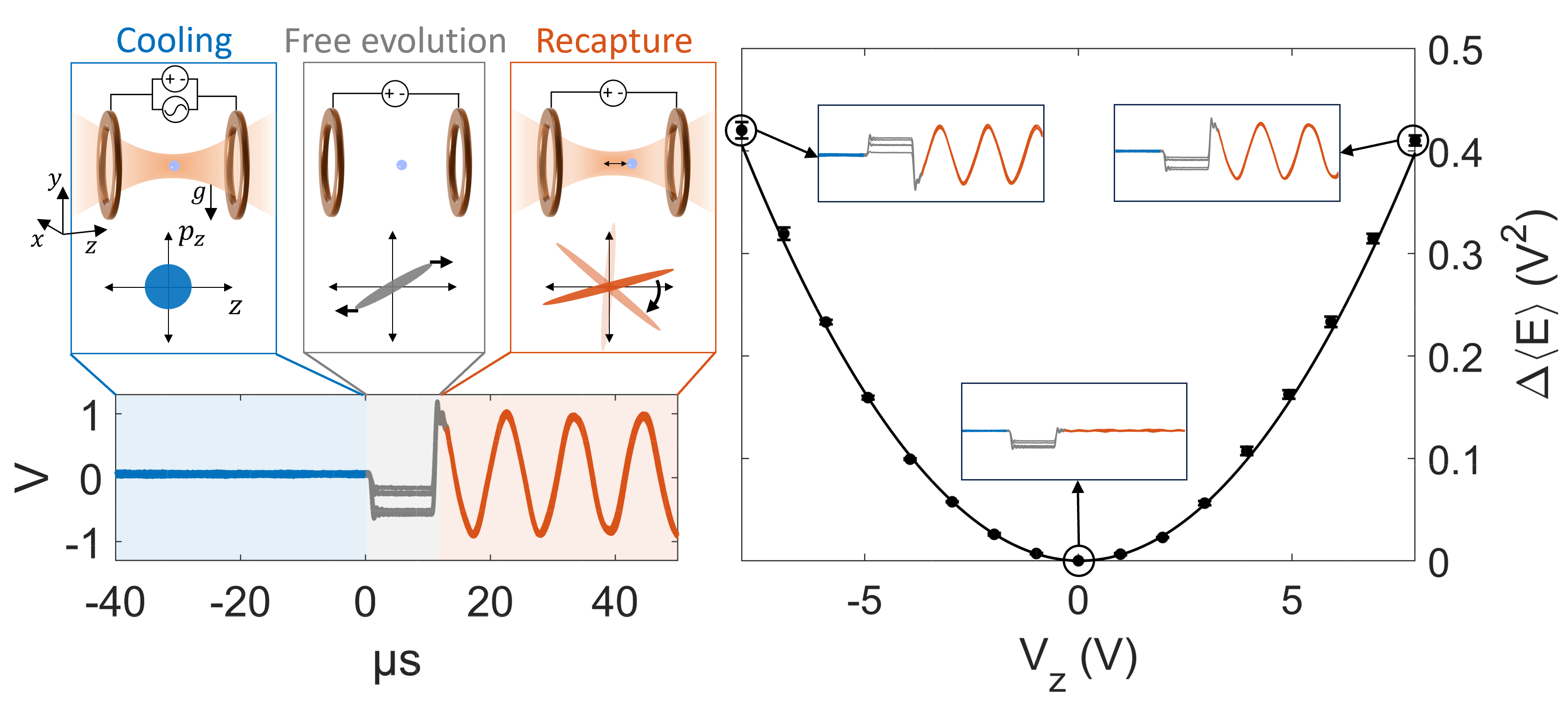}
    \caption{The electrostatic compensation procedure for the axial direction. \textbf{Left:} The three panels show the experimental sequence, with only the axial electrode pair illustrated: cooling (trap + feedback + DC bias), free evolution (DC bias) and recapture (trap + DC bias). The corresponding phase space distribution at each step is also illustrated. \textbf{Right:} A measurement showing the increase in mean energy over the sequence as a function of the applied DC bias in the axial ($z$) direction, with five repetions per voltage. A parabolic fit (black line) indicates that the optimal compensation voltage is 0.04\,V in this example. Insets are sample timetraces at the various biases, with blue indicating the time before free evolution, red the time after recapture, and gray the free evolution plus a buffer to account for the detector response.}
    \label{fig:fig1}
\end{figure}

%%% Experiment description %%
The experiment consists of an optical tweezer in ultra-high vacuum surrounded by three orthogonal pairs of copper ring electrodes (see Supplementary Material).
We use a Coherent Mephisto 1064\,nm laser with typical trapping powers of 0.6\,W and Asphericon 0.8 NA trapping and collection lenses. 
Silica particles with a mean diameter of 156\,nm (MicroParticles Gmbh) are loaded into the trap via nebulization.
Back-scattered light is separated from the trapping light with a Faraday rotator, confocally coupled to a fiber and measured in a phase-locked homodyne scheme~\cite{Magrini2021} to cool the axial motion to about 100 phonons.
Split detection of the forward-scattered light is used for linear cooling of the radial modes, enabling stable trapping at low pressures.

The feedback signal is generated from phase-locked loops implemented on a combination of Red Pitaya FPGAs (STEMLab 125-14) and a Zurich Instruments HF2LI lock-in amplifier (for more robust cooling of the radial modes) and then sent to the appropriate pair of electrodes. 
The feedback signals are combined with the DC compensation voltages via a bias tee and then applied to the electrodes.
The DC voltages are supplied by three Stanford Research Systems DC205 voltage supplies, chosen for their long-term ppm-level stability.
In the radial directions, the feedback and DC compensation voltages are further amplified with Falco WMA-100A amplifiers.

%%% 1d compensation %%%
The procedure for 1D force compensation is illustrated in Fig.~\ref{fig:fig1}.
At a given DC bias, we simultaneously switch off the tweezer and feedback for a time $\tau$, after which the tweezer is turned back on and the particle recaptured in the trap.
The feedback is re-enabled after an additional delay (typically 6\,ms) such that feedback is not active for any of the post-recapture data.
Plotting the mean energy after recapture over a range of compensation voltages (see Supplementary Material for details on how the mean energy is extracted) yields a parabola, the minimum of which, $V_{\text{opt}}$, corresponds to the total background force which needs to be compensated along the scanned axis.
Typical $V_{\text{opt}}$ for the axial ($z$) direction are in the $\pm1$\,V range; for the example shown in Fig.~\ref{fig:fig1} it is close to 0\,V.

The back-scattered homodyne detection is used to determine $V_{\text{opt}}$, since it provides the most sensitive detection for all directions.
This procedure is most sensitive for the axial direction, for which the electrode geometry yields an approximately 100 times stronger Newton-to-Volt transduction coefficient compared to the radial directions ($10^{-16}$ N/V compared to $10^{-18}$ N/V), and for which the starting occupancy is lowest.
Estimates for the required compensation accuracy along each axis are provided in the Supplementary Material.

For electrostatic stray fields, the optimal compensation voltages are independent of the particle charge state, and therefore provide an in-situ measurement of the environmental charge along a given axis.
By performing 1D compensation scans along each axis, we determined that the stray fields contribute comparably in all axes, typically of order the gravitational force, which means that 3D compensation is necessary to minimize displacement during the free evolution.

\begin{figure}
    \centering
    \includegraphics[width=0.8\linewidth]{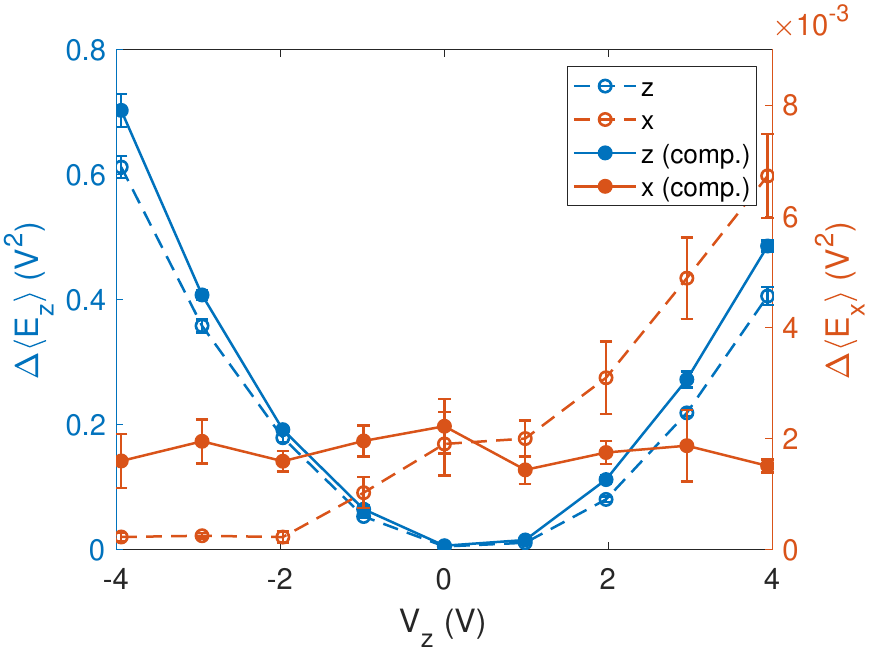}
    \caption{The effect of force cross-talk on axial compensation scans. We compute power spectral densities using 5\,ms of data before and after the 10\,µs free evolutions to isolate the response of the individual motional modes. The $z$ mode shows the expected parabolic behavior (blue), while the $x$ mode shows a clear dependence on $V_z$ due to cross-talk (red dashed line). When the voltages are instead scanned in a proportion given by the third column of $C^{-1}$ (solid lines), the cross-talk into $x$ is significantly reduced. Each point contains five repetitions and error bars are standard errors of the mean.}
    \label{fig:cross_talk}
\end{figure}

Extension of this technique to 3D by simply iterating the procedure for the other axes is complicated by the presence of cross-talk between the electrode axes as measured in the trap basis. 
This is a consequence of a mismatch between the electrode ($x',y',z'$) and trap ($x,y,z$) coordinate systems, due to e.g. imperfections in electrode construction or the polarization of the trapping beam.
A voltage applied to a given electrode therefore yields a force $\mathbf{F_{\text{trap}}}$ along all three trap axes according to the projection matrix $C$,
\begin{equation}
    \mathbf{F_{\text{trap}}} = 
    \begin{pmatrix} C_{x'x} & C_{y'x} & C_{z'x} \\ C_{x'y} & C_{y'y} & C_{z'y} \\ C_{x'z} & C_{y'z} & C_{z'z} \end{pmatrix}
    \begin{pmatrix} V_{x'} \\ V_{y'} \\ V_{z'} \end{pmatrix}
    \coloneq C\mathbf{V}.
\end{equation}
To avoid confusion, we will continue to refer to "$x$/$y$/$z$" electrodes without primes for the electrodes used for cooling of the $x$/$y$/$z$ motional modes.  

$C^{-1}$ represents a geometric correction factor: its columns contain the combination of electrode voltages which yield a force along one trap axis at a time.
The entries of $C^{-1}$ could be determined by measuring the response to an electronic harmonic drive applied to the various axes in independent detections~\cite{Zhu2023}. 
Alternatively, one could design the electrodes to geometrically minimize the contribution of cross-talk, as has recently been demonstrated in the context of feedback cooling~\cite{gosling2025}.

Instead, we developed a method which can be used with arbitrary 3D electrode configurations, is robust to unstable mechanical frequencies and is calibration-free.
Just like for the harmonic drive method, it requires that each detection is primarily sensitive to motion along one trap axis. 
We determine the magnitude of the various cross-talk terms by cooling a given mode with the appropriate set of electrodes (e.g., the $z$ mode with the $z$ electrodes with a voltage $V_{z'}^z$) and then measuring the voltage required to "cross-cool" the same mode to the same amplitude via an orthogonal set of electrodes (e.g. via the cross-talk of the $x$ electrodes, $V_{x'}^z$).
The feedback signal for cooling is obtained via a phase-locked loop locked to the mode of interest, and the mode amplitude is monitored by demodulating this phase-locked loop with the detection signal. 

This procedure yields equations of the form $C_{x'i}V_{x'}^i = C_{y'i}V_{y'}^i = C_{z'i}V_{z'}^i$ for each trap axis $i \in \{x, y, z\}$, which are then used to determine the entries of $C^{-1}$ up to three scale factors given by $C_{i'i}$.
The resulting columns of $C^{-1}$ then define the proportion of voltages needed to apply a force along a given trap axis, thus allowing us to perform compensation scans iteratively for each axis.

An example of a cross-talk-corrected compensation scan is shown in Fig.~\ref{fig:cross_talk}.
Integrating the area under the various motional peaks in the spectrum before and after the free evolution allows us to estimate the effect of a given set of voltages on each axis individually.
The flattening of the $x$ mode response as a function of $V_z$ shows that the contribution of the $C_{z'x}V_{z'}$ cross-talk term has been minimized. 

%%% Record drops and discussion %%%
\begin{figure}
    \centering
    \includegraphics[width=1.0\linewidth]{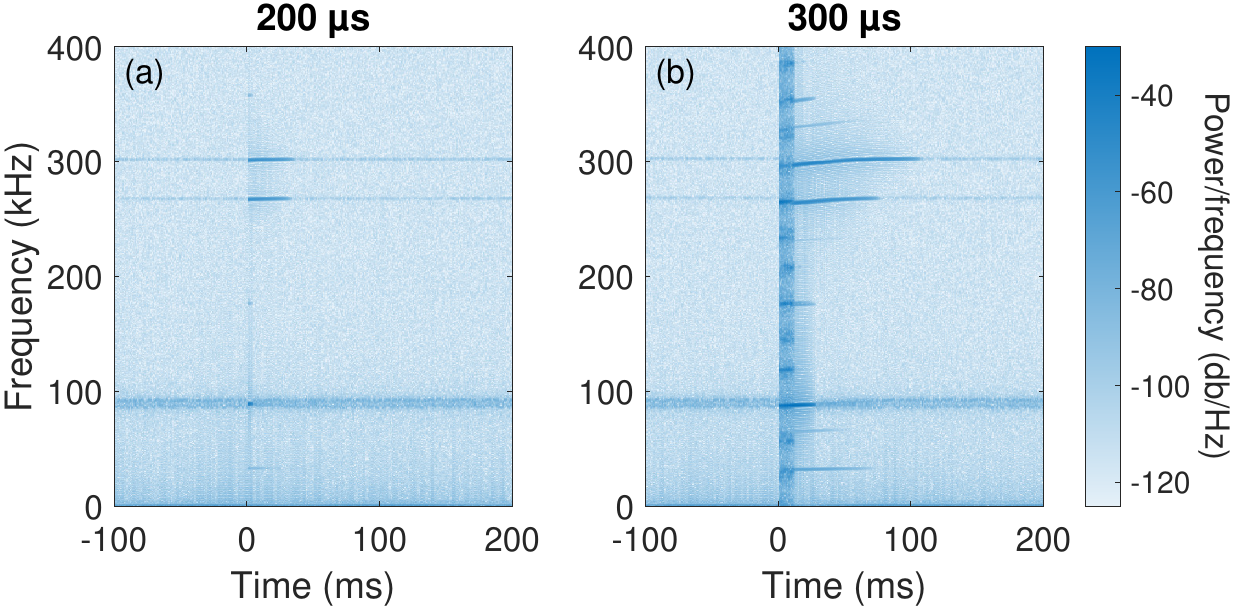}
    \caption{Spectrogram of a single 200\,µs (a) and a single 300\,µs (b) free evolution followed by recapture as measured in the back-scattered homodyne detection. The trap and feedback are switched off at $t=0$ and re-enabled simultaneously after the free expansion. Excitation and re-cooling of the motional modes $\Omega_{x,y,z}/(2\pi) = [302, 268, 92]$\,kHz is visible upon recapture. In the 300\,µs spectrogram, a several kHz frequency drop and slow recovery is also visible, likely due to imperfect radial compensation and a resulting exploration of trap nonlinearites (see Supplementary Material).}
    \label{fig:spectro}
\end{figure}

The optimal compensation voltages in all trap axes can now be determined by performing cross-talk-free compensation scans of each axis iteratively and with increasing $\tau$ for improved accuracy.
Using the resulting compensation voltages, free evolutions of several 100\,µs are possible with minimal excitation of the motional modes, as shown in Fig.~\ref{fig:spectro}.
For even longer durations, we observe stronger motional excitation and a corresponding drop in the frequencies upon recapture, which we attribute primarily to imperfect radial compensation and a corresponding probing of trap nonlinearities~\cite{Gieseler2013}.
The radial frequency drop limits the maximum free evolution duration because of limitations of the radial feedback, resulting in loss of radial cooling and eventual particle loss. 
The higher radial occupancy in our experiment is also detrimental because higher center-of-mass temperatures yield a larger mean-squared displacement due to Brownian motion, which cannot be compensated with a DC force.
For coherent free evolutions on the millisecond timescale, lower pressure and lower starting occupancies in all axes will therefore be crucial. 

We can quantify how well DC forces are compensated by measuring the mean energy as a function of $\tau$. 
In the presence of a constant axial force $F_z$, the axial mean energy increases as~\cite{Hebestreit2018}
\begin{equation}
\label{eq:energy}
    \langle E_z \rangle = E_z(0)\left[1 + \frac{\Omega^2\tau^2}{2}\right] + F_z^2\frac{\tau^2}{2m}\left[1+\frac{\Omega^2\tau^2}{4}\right]
\end{equation}
with $E_z(0)$ the energy of the mode before the free evolution. 
The $F^2$ scaling of the mean energy is observed during the voltage compensation scans, where $\tau$ is held constant, while the scaling with $\tau^4$ at large $\tau$ is shown in the Supplementary Material. 

\begin{figure}
    \centering
    \includegraphics[width=0.8\linewidth]{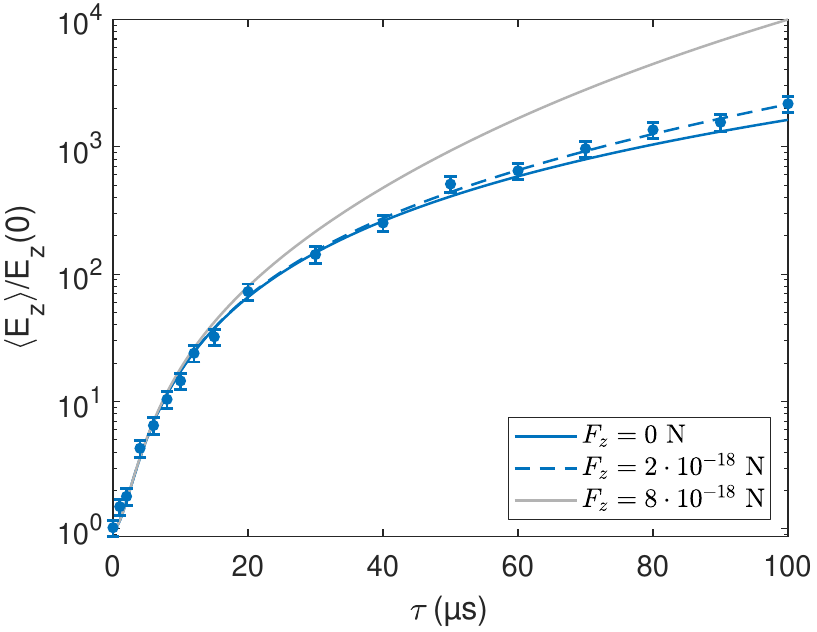}
    \caption{The relative mean energy increase of the axial mode as a function of $\tau$ at pressures below $8\cdot10^{-8}$\,mbar. For force-free evolution, this should evolve as $1+\Omega^2\tau^2/2$ (blue line). The data is consistent with a $2\cdot 10^{-18}$\,N residual force (blue dashed line, corresponding to a 2.5\,nm mean displacement over 100\,µs) but strongly excludes larger forces. For example, the gray line corresponds to a force which would yield mean displacements of 10\,nm over 100\,µs. The mean energy is extracted from the squared amplitudes of a sine fit, with each data point corresponding to 150 repetitions and error bars the standard error of the mean.}
    \label{fig:meanE}
\end{figure}

The scaling of the mean energy with $1 + \Omega^2\tau^2/2$ can be seen in Fig.~\ref{fig:meanE}, where the measured relative energy increase of the axial mode is shown as a function of $\tau$, with 3D compensation voltages applied.
The mean energy is extracted as the squared amplitudes of sinusoidal fits to the unfiltered oscillations (see Supplementary Material for details).
The data shows good agreement with the expected evolution in the absence of an external force, demonstrating the efficacy of the DC compensation.
In particular, the mean displacement along $z$ is confined to less than 10\,nm for free evolutions up to 100\,µs.

In addition to mean energy, one can estimate the evolution of the state variance by taking the variance of the fitted trajectories after recapture. 
The evolution of the full covariance matrix is calculated via the differential Lyapunov equation~\cite{Genoni2016}, which also allows for straightforward implementation of momentum diffusion during the free evolution and trap evolution, dominated by gas collisions and recoil, respectively.
The axial ($z$) position variance (the first entry of the covariance matrix) after a free evolution of time $\tau$ and subsequent evolution in a harmonic for a time $t$, in the absence of diffusion, is given by 
\begin{equation}
    \sigma^2_z(t) =  \sigma^2_z(0)\left[1 + \Omega \tau  \sin(2\Omega t)\right] + \sigma^2_p(0)\frac{\tau^2}{m^2} \cos^2\Omega t,
\end{equation}
with $\sigma^2_z(0) = z_{\text{zp}}^2(2\bar{n}+1)$ and $\sigma^2_p(0) = z_{\text{zp}}^2(2\bar{n}+1)m^2\Omega^2$. 

\begin{figure}
    \centering
    \includegraphics[width=0.8\linewidth]{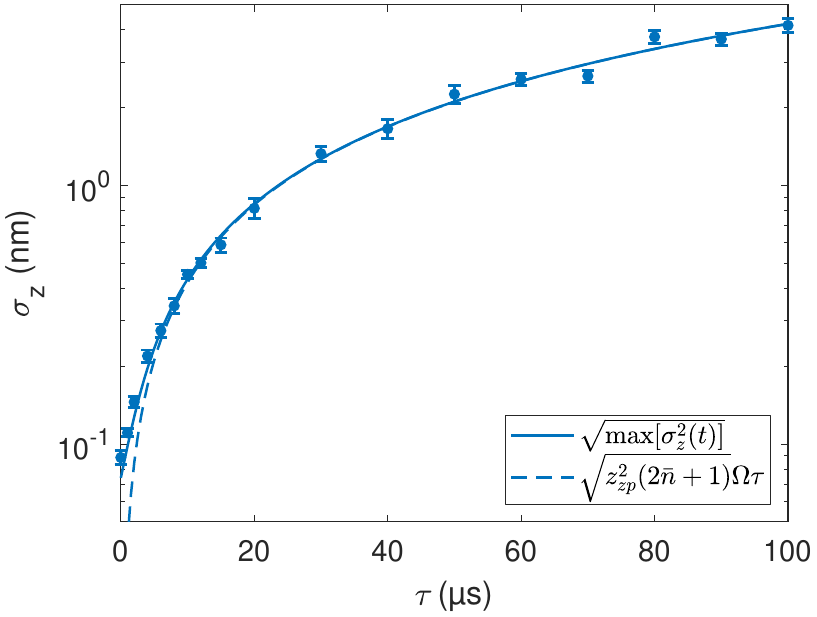}
    \caption{Growth of the position standard deviation of the axial mode as a function of $\tau$ for the same dataset as Fig.~\ref{fig:meanE}. The state variance evolution is extracted by analyzing the statistics of the sine-fitted trajectories upon recapture. We plot the maximum of the standard deviation as a function of $\tau$, which grows as $\Omega\tau$ for large $\tau$ (blue dashed line), with the full theory the blue solid line (the effect of gas and recoil is negligible). The occupancy $\bar{n}=117$ used in the theory curve is estimated by fitting to $z_{\text{zp}}^2(2\bar{n}+1)(1+\Omega^2\tau^2)$, which agrees to within 10\% with the occupancy estimated via integration of the calibrated spectrum. Each point contains 150 repetitions, with error bars estimated via binning (see Supplementary Material).}
    \label{fig:sigma}
\end{figure}

The position variance $\sigma^2_z(t)$ oscillates at $2\Omega$ as the sheared distribution rotates in the harmonic potential.
For $\tau\gg\Omega^{-1}$, $\max[\sigma^2_z(t)] \approx \sigma^2_p(0)\tau^2/m^2$, i.e., the maximum position variance at large $\tau$ is a time-of-flight measurement of the momentum variance of the state prior to the free expansion.
We plot the maximum standard deviation of the fitted trajectories in Fig.~\ref{fig:sigma}, showing excellent agreement with the full theory model and the $\Omega^2\tau^2$ approximation for large $\tau$.

For 100\,µs free evolutions, the observed state extent of 4\,nm corresponds to a 56-fold expansion of the initial 74\,pm thermal state.
The agreement with the model also rules out a significant contribution of fluctuating forces during the free evolution, since this would contribute an increase in variance not present in the model.

%%% Coherences
Expansion for nanoparticle interferometry protocols should also be coherent, as a reduction in purity would affect the visibility of any interference fringes. 
To investigate the role of voltage noise of our DC compensation supplies, we performed reheating measurements with various DC bias voltages applied to the electrodes at pressures below 6$\cdot10^{-8}$\,mbar (see Supplementary Material).
We observe an increase in heating rate only for axial voltages above 50\,V , far beyond typical compensation voltages ($\approx$1\,V), implying that the voltage noise is not a significant heating mechanism at these pressures.
Finally, we also perform a recompression protocol (see Supplementary Material) and find that the state recompresses to its initial extent as expected, implying no significant loss of purity during short (5\,µs) free evolutions with 3D compensation. 

%%% Charge control %%%
A charged nanoparticle and a stably (dis)charged environment are essential for the electrostatic force compensation we implement.
There are a number of existing techniques for nanoparticle charge control, including high-voltage plasma discharge~\cite{Frimmer2017} and UV illumination~\cite{Moore2014}. 
However, by employing the particle as a force sensor, we determined that the plasma discharge technique significantly charges the environment, based on an order-of-magnitude increase of the required axial compensation voltage. 
This is likely due to the deposition of charges on the trapping/collections lenses as well as the PEEK electrode holder, since these are the closest dielectric surfaces to the particle. 
We therefore typically avoid plasma discharges and work with natively charged particles (with typically tens of elementary charges).
If particle charge control is required, other techniques, such as focused UV illumination, could provide a means to modify the particle charge with a more moderate effect on the environment. 

Reducing the environmental charge state is desirable because the compensation voltage supplies have a finite range, and because higher compensation voltages are associated with increased voltage noise.  
For this purpose, we developed a method to control the environmental charge independently of the particle's charge state and at low pressure, which avoids uncontrolled triboelectric charging during pumping.
By passing current through a  tungsten filament installed directly in the vacuum chamber, we generate a flux of low-energy electrons via thermionic emission. 
By iteratively "flashing" the filament with the axial electrodes biased at different values, we could guide the electrons to various dielectric surfaces within the chamber to achieve the desired net environmental charge, as shown in the Supplemental Material.
However, we found that the resulting environmental charge state slowly stabilized over hours (see Supplemental Material), which was detrimental for long measurements.
The method is still valuable as a vacuum-compatible approach for environmental charge control, especially for measurements insensitive to slow drifts of the background fields, or if the requisite time is allowed for stabilization. 

%%% Summary %%%
In conclusion, we have demonstrated a simple but effective procedure for implementing 3D electrostatic force compensation of a charged nanoparticle.
Our technique enables stable free evolutions of 100\,µs duration without significant mean displacement and an increase in position variance consistent with coherent dynamics.
By implementing this at lower pressures and starting with a purer initial state, we expect millisecond-long coherent free expansions to become possible, enabling the wavepacket expansion required for preparation of squeezed states, nanoparticle interferometry and a sensitive time-of-flight technique for the measurement of interference fringes in momentum space.
We have also demonstrated how to use the nanoparticle as a sensitive environmental DC force sensor, which could be of interest for experiments that require well-calibrated and stable environments\cite{Moore2021}. 
We anticipate that the methods we have developed to enable long, coherent free evolutions of charged nanoparticles will provide an essential tool for further advancing levitated experiments into the quantum regime. 

\section*{Supplementary Material}
See the Supplementary Material for additional information on experimental details and electrode geometry, the cross-cooling method, compensation scan scalings with expansion time, charge characterization and control, reheating/recompression measurements, estimations of the required compensation voltage accuracy and analysis details.

\begin{acknowledgments}
We thank the Q-Xtreme synergy group for helpful discussions, Martin Duchaň for his help with characterizing the electrode cross-talk and Rachel Goodman for her early work on the setup.
This project was supported by the European Research Council under the European Union's Horizon 2020 research and innovation program (ERC Synergy Q-Xtreme, Grant No. 951234).
This research was funded in whole or in part by the Austrian Science Fund (FWF) [10.55776/COE1 and 10.55776/PAT9140723].
For Open Access purposes, the author has applied a CC BY public copyright license to any author accepted manuscript version arising from this submission.

\end{acknowledgments}

\bibliography{references}

@article{Bateman2014,
  title = {Near-field interferometry of a free-falling nanoparticle from a point-like source},
  volume = {5},
  ISSN = {2041-1723},
  url = {http://dx.doi.org/10.1038/ncomms5788},
  DOI = {10.1038/ncomms5788},
  number = {1},
  journal = {Nature Communications},
  publisher = {Springer Science and Business Media LLC},
  author = {Bateman,  James and Nimmrichter,  Stefan and Hornberger,  Klaus and Ulbricht,  Hendrik},
  year = {2014},
  month = Sep 
}

@article{Bonvin2024,
  title = {State Expansion of a Levitated Nanoparticle in a Dark Harmonic Potential},
  author = {Bonvin, Eric and Devaud, Louisiane and Rossi, Massimiliano and Militaru, Andrei and Dania, Lorenzo and Bykov, Dmitry S. and Romero-Isart, Oriol and Northup, Tracy E. and Novotny, Lukas and Frimmer, Martin},
  journal = {Phys. Rev. Lett.},
  volume = {132},
  issue = {25},
  pages = {253602},
  numpages = {7},
  year = {2024},
  month = {Jun},
  publisher = {American Physical Society},
  doi = {10.1103/PhysRevLett.132.253602},
  url = {https://link.aps.org/doi/10.1103/PhysRevLett.132.253602}
}

@article{Brown2023,
  title     = "Time-of-flight quantum tomography of an atom in an optical tweezer",
  author    = "Brown, M O and Muleady, S R and Dworschack, W J and Lewis-Swan,
               R J and Rey, A M and Romero-Isart, O and Regal, C A",
  journal   = "Nat. Phys.",
  publisher = "Springer Science and Business Media LLC",
  volume    =  19,
  number    =  4,
  pages     = "569--573",
  month     =  apr,
  year      =  2023,
  copyright = "https://www.springernature.com/gp/researchers/text-and-data-mining",
}

@article{Carney2021,
doi = {10.1088/2058-9565/abcfcd},
url = {https://dx.doi.org/10.1088/2058-9565/abcfcd},
year = {2021},
month = {jan},
publisher = {IOP Publishing},
volume = {6},
number = {2},
pages = {024002},
author = {Carney, D and Krnjaic, G and Moore, D C and Regal, C A and Afek, G and Bhave, S and Brubaker, B and Corbitt, T and Cripe, J and Crisosto, N and Geraci, A and Ghosh, S and Harris, J G E and Hook, A and Kolb, E W and Kunjummen, J and Lang, R F and Li, T and Lin, T and Liu, Z and Lykken, J and Magrini, L and Manley, J and Matsumoto, N and Monte, A and Monteiro, F and Purdy, T and Riedel, C J and Singh, R and Singh, S and Sinha, K and Taylor, J M and Qin, J and Wilson, D J and Zhao, Y},
title = {Mechanical quantum sensing in the search for dark matter},
journal = {Quantum Science and Technology}
}

@article{Dago24,
author = {Salamb\^{o} Dago and J. Rieser and M. A. Ciampini and V. Mlyn\'{a}\v{r} and A. Kugi and M. Aspelmeyer and A. Deutschmann-Olek and N. Kiesel},
journal = {Opt. Express},
number = {25},
pages = {45133--45141},
publisher = {Optica Publishing Group},
title = {Stabilizing nanoparticles in the intensity minimum: feedback levitation on an inverted potential},
volume = {32},
month = {Dec},
year = {2024},
url = {https://opg.optica.org/oe/abstract.cfm?URI=oe-32-25-45133},
doi = {10.1364/OE.541267},
}

@article{Duchan2025,
  title = {Nanomechanical state amplifier based on optical inverted pendulum},
  volume = {8},
  ISSN = {2399-3650},
  url = {http://dx.doi.org/10.1038/s42005-025-02193-z},
  DOI = {10.1038/s42005-025-02193-z},
  number = {1},
  journal = {Communications Physics},
  publisher = {Springer Science and Business Media LLC},
  author = {Duchaň,  Martin and Šiler,  Martin and Jákl,  Petr and Brzobohatý,  Oto and Rakhubovsky,  Andrey and Filip,  Radim and Zemánek,  Pavel},
  year = {2025},
  month = jul 
}

@article{Frimmer2017,
  title = {Controlling the net charge on a nanoparticle optically levitated in vacuum},
  author = {Frimmer, Martin and Luszcz, Karol and Ferreiro, Sandra and Jain, Vijay and Hebestreit, Erik and Novotny, Lukas},
  journal = {Phys. Rev. A},
  volume = {95},
  issue = {6},
  pages = {061801},
  numpages = {4},
  year = {2017},
  month = {Jun},
  publisher = {American Physical Society},
  doi = {10.1103/PhysRevA.95.061801},
  url = {https://link.aps.org/doi/10.1103/PhysRevA.95.061801}
}

@article{Gasbarri2021,
  title = {Testing the foundation of quantum physics in space via Interferometric and non-interferometric experiments with mesoscopic nanoparticles},
  volume = {4},
  ISSN = {2399-3650},
  url = {http://dx.doi.org/10.1038/s42005-021-00656-7},
  DOI = {10.1038/s42005-021-00656-7},
  number = {1},
  journal = {Communications Physics},
  publisher = {Springer Science and Business Media LLC},
  author = {Gasbarri,  Giulio and Belenchia,  Alessio and Carlesso,  Matteo and Donadi,  Sandro and Bassi,  Angelo and Kaltenbaek,  Rainer and Paternostro,  Mauro and Ulbricht,  Hendrik},
  year = {2021},
  month = jul 
}

@article{Genoni2016,
author = {Marco G. Genoni and Ludovico Lami and Alessio Serafini},
title = {Conditional and unconditional Gaussian quantum dynamics},
journal = {Contemporary Physics},
volume = {57},
number = {3},
pages = {331--349},
year = {2016},
publisher = {Taylor \& Francis},
doi = {10.1080/00107514.2015.1125624},
URL = {https://doi.org/10.1080/00107514.2015.1125624},
eprint = {https://doi.org/10.1080/00107514.2015.1125624}
}

@article{Gieseler2013,
  title = {Thermal nonlinearities in a nanomechanical oscillator},
  volume = {9},
  ISSN = {1745-2481},
  url = {http://dx.doi.org/10.1038/nphys2798},
  DOI = {10.1038/nphys2798},
  number = {12},
  journal = {Nature Physics},
  publisher = {Springer Science and Business Media LLC},
  author = {Gieseler,  Jan and Novotny,  Lukas and Quidant,  Romain},
  year = {2013},
  month = nov,
  pages = {806–810}
}

@misc{gosling2025,
      title={Feedback cooling scheme for an optically levitated oscillator with controlled cross-talk}, 
      author={J. M. H. Gosling and A. Pontin and F. Alder and M. Rademacher and T. S. Monteiro and P. F. Barker},
      year={2025},
      eprint={2506.17172},
      archivePrefix={arXiv},
      primaryClass={physics.optics},
      url={https://arxiv.org/abs/2506.17172}, 
}

@article{Hebestreit2018,
  title = {Sensing Static Forces with Free-Falling Nanoparticles},
  author = {Hebestreit, Erik and Frimmer, Martin and Reimann, Ren\'e and Novotny, Lukas},
  journal = {Phys. Rev. Lett.},
  volume = {121},
  issue = {6},
  pages = {063602},
  numpages = {5},
  year = {2018},
  month = {Aug},
  publisher = {American Physical Society},
  doi = {10.1103/PhysRevLett.121.063602},
  url = {https://link.aps.org/doi/10.1103/PhysRevLett.121.063602}
}

@article{Hebestreit2018_calib,
    author = {Hebestreit, Erik and Frimmer, Martin and Reimann, René and Dellago, Christoph and Ricci, Francesco and Novotny, Lukas},
    title = {Calibration and energy measurement of optically levitated nanoparticle sensors},
    journal = {Review of Scientific Instruments},
    volume = {89},
    number = {3},
    pages = {033111},
    year = {2018},
    month = {03},
    issn = {0034-6748},
    doi = {10.1063/1.5017119},
    url = {https://doi.org/10.1063/1.5017119},
    eprint = {https://pubs.aip.org/aip/rsi/article-pdf/doi/10.1063/1.5017119/15951914/033111\_1\_online.pdf},
}

@article{Kamba2023,
  title = {Revealing the Velocity Uncertainties of a Levitated Particle in the Quantum Ground State},
  author = {Kamba, M. and Aikawa, K.},
  journal = {Phys. Rev. Lett.},
  volume = {131},
  issue = {18},
  pages = {183602},
  numpages = {6},
  year = {2023},
  month = {Oct},
  publisher = {American Physical Society},
  doi = {10.1103/PhysRevLett.131.183602},
  url = {https://link.aps.org/doi/10.1103/PhysRevLett.131.183602}
}

@misc{kamba2025,
      title={Quantum squeezing of a levitated nanomechanical oscillator}, 
      author={M. Kamba and N. Hara and K. Aikawa},
      year={2025},
      eprint={2504.17944},
      archivePrefix={arXiv},
      primaryClass={quant-ph},
      url={https://arxiv.org/abs/2504.17944}, 
}

@article{Kialka2022,
    author = {Kiałka, Filip and Fein, Yaakov Y. and Pedalino, Sebastian and Gerlich, Stefan and Arndt, Markus},
    title = "{A roadmap for universal high-mass matter-wave interferometry}",
    journal = {AVS Quantum Science},
    volume = {4},
    number = {2},
    pages = {020502},
    year = {2022},
    month = {04},
    issn = {2639-0213},
    doi = {10.1116/5.0080940},
    url = {https://doi.org/10.1116/5.0080940},
    eprint = {https://pubs.aip.org/avs/aqs/article-pdf/doi/10.1116/5.0080940/19957266/020502\_1\_5.0080940.pdf},
}

@article{levitoreview,
    author = {C. Gonzalez-Ballestero  and M. Aspelmeyer  and L. Novotny  and R. Quidant  and O. Romero-Isart },
    title = {Levitodynamics: Levitation and control of microscopic objects in vacuum},
    journal = {Science},
    volume = {374},
    number = {6564},
    pages = {eabg3027},
    year = {2021},
    doi = {10.1126/science.abg3027}
}

@article{Magrini2021,
   author = {Lorenzo Magrini and Philipp Rosenzweig and Constanze Bach and Andreas Deutschmann-Olek and Sebastian G. Hofer and Sungkun Hong and Nikolai Kiesel and Andreas Kugi and Markus Aspelmeyer},
   doi = {10.1038/s41586-021-03602-3},
   issn = {14764687},
   issue = {7867},
   journal = {Nature},
   month = {7},
   pages = {373-377},
   pmid = {34262213},
   publisher = {Nature Research},
   title = {Real-time optimal quantum control of mechanical motion at room temperature},
   volume = {595},
   year = {2021},
}

@misc{mattana2025,
      title={Trap-to-trap free falls with an optically levitated nanoparticle}, 
      author={M. Luisa Mattana and Nicola Carlon Zambon and Massimiliano Rossi and Eric Bonvin and Louisiane Devaud and Martin Frimmer and Lukas Novotny},
      year={2025},
      eprint={2507.12995},
      archivePrefix={arXiv},
      primaryClass={quant-ph},
      url={https://arxiv.org/abs/2507.12995}, 
}

@article{Millen2020,
doi = {10.1088/1361-6633/ab6100},
url = {https://dx.doi.org/10.1088/1361-6633/ab6100},
year = {2020},
month = {jan},
publisher = {IOP Publishing},
volume = {83},
number = {2},
pages = {026401},
author = {Millen, James and Monteiro, Tania S and Pettit, Robert and Vamivakas, A Nick},
title = {Optomechanics with levitated particles},
journal = {Reports on Progress in Physics}
}

@article{Moore2014,
  title = {Search for Millicharged Particles Using Optically Levitated Microspheres},
  author = {Moore, David C. and Rider, Alexander D. and Gratta, Giorgio},
  journal = {Phys. Rev. Lett.},
  volume = {113},
  issue = {25},
  pages = {251801},
  numpages = {5},
  year = {2014},
  month = {Dec},
  publisher = {American Physical Society},
  doi = {10.1103/PhysRevLett.113.251801},
  url = {https://link.aps.org/doi/10.1103/PhysRevLett.113.251801}
}

@article{Moore2021,
doi = {10.1088/2058-9565/abcf8a},
url = {https://dx.doi.org/10.1088/2058-9565/abcf8a},
year = {2021},
month = {Jan},
publisher = {IOP Publishing},
volume = {6},
number = {1},
pages = {014008},
author = {David C Moore and Andrew A Geraci},
title = {Searching for new physics using optically levitated sensors},
journal = {Quantum Science and Technology}
}

@article{Neumeier2024,
author = {Lukas Neumeier  and Mario A. Ciampini  and Oriol Romero-Isart  and Markus Aspelmeyer  and Nikolai Kiesel },
title = {Fast quantum interference of a nanoparticle via optical potential control},
journal = {Proceedings of the National Academy of Sciences},
volume = {121},
number = {4},
pages = {e2306953121},
year = {2024},
doi = {10.1073/pnas.2306953121},
URL = {https://www.pnas.org/doi/abs/10.1073/pnas.2306953121},
eprint = {https://www.pnas.org/doi/pdf/10.1073/pnas.2306953121}}

@article{Pino2018,
doi = {10.1088/2058-9565/aa9d15},
url = {https://dx.doi.org/10.1088/2058-9565/aa9d15},
year = {2018},
month = {jan},
publisher = {IOP Publishing},
volume = {3},
number = {2},
pages = {025001},
author = {Pino, H and Prat-Camps, J and Sinha, K and Venkatesh, B Prasanna and Romero-Isart, O},
title = {On-chip quantum interference of a superconducting microsphere},
journal = {Quantum Science and Technology}
}

@misc{prakash2025,
      title={Release and Recapture of Silica Nanoparticles from an Optical Trap in Weightlessness}, 
      author={Govindarajan Prakash and Sven Herrmann and Ralf B. Bergmann and Christian Vogt},
      year={2025},
      eprint={2509.08666},
      archivePrefix={arXiv},
      primaryClass={physics.optics},
      url={https://arxiv.org/abs/2509.08666}, 
}

@article{Roda_Llordes2024,
  title = {Macroscopic Quantum Superpositions via Dynamics in a Wide Double-Well Potential},
  author = {Roda-Llordes, M. and Riera-Campeny, A. and Candoli, D. and Grochowski, P. T. and Romero-Isart, O.},
  journal = {Phys. Rev. Lett.},
  volume = {132},
  issue = {2},
  pages = {023601},
  numpages = {7},
  year = {2024},
  month = {Jan},
  publisher = {American Physical Society},
  doi = {10.1103/PhysRevLett.132.023601},
  url = {https://link.aps.org/doi/10.1103/PhysRevLett.132.023601}
}

@article{Romero-Isart2017,
doi = {10.1088/1367-2630/aa99bf},
url = {https://dx.doi.org/10.1088/1367-2630/aa99bf},
year = {2017},
month = {dec},
publisher = {IOP Publishing},
volume = {19},
number = {12},
pages = {123029},
author = {Oriol Romero-Isart},
title = {Coherent inflation for large quantum superpositions of levitated microspheres},
journal = {New Journal of Physics}
}

@misc{rossi2024,
      title={Quantum Delocalization of a Levitated Nanoparticle}, 
      author={Massimiliano Rossi and Andrei Militaru and Nicola Carlon Zambon and Andreu Riera-Campeny and Oriol Romero-Isart and Martin Frimmer and Lukas Novotny},
      year={2024},
      eprint={2408.01264},
      archivePrefix={arXiv},
      primaryClass={quant-ph},
      url={https://arxiv.org/abs/2408.01264}, 
}

@misc{tomassi2025,
      title={Accelerated State Expansion of a Nanoparticle in a Dark Inverted Potential}, 
      author={Gregoire F. M. Tomassi and Daniel Veldhuizen and Bruno Melo and Davide Candoli and Andreu Riera-Campeny and Oriol Romero-Isart and Nadine Meyer and Romain Quidant},
      year={2025},
      eprint={2503.20707},
      archivePrefix={arXiv},
      primaryClass={quant-ph},
      url={https://arxiv.org/abs/2503.20707}, 
}

@article{Ugolini2011,
    author = {Ugolini, D. and Funk, Q. and Amen, T.},
    title = "{Note: Discharging fused silica test masses with ionized nitrogen}",
    journal = {Review of Scientific Instruments},
    volume = {82},
    number = {4},
    pages = {046108},
    year = {2011},
    month = {04},
    issn = {0034-6748},
    doi = {10.1063/1.3579500},
    url = {https://doi.org/10.1063/1.3579500},
    eprint = {https://pubs.aip.org/aip/rsi/article-pdf/doi/10.1063/1.3579500/13461931/046108\_1\_online.pdf},
}

@article{Weiss2021,
archivePrefix = {arXiv},
arxivId = {2012.12260},
author = {Weiss, T. and Roda-Llordes, M. and Torrontegui, E. and Aspelmeyer, M. and Romero-Isart, O.},
doi = {10.1103/PhysRevLett.127.023601},
eprint = {2012.12260},
issn = {0031-9007},
journal = {Physical Review Letters},
month = {jul},
number = {2},
pages = {023601},
title = {{Large Quantum Delocalization of a Levitated Nanoparticle Using Optimal Control: Applications for Force Sensing and Entangling via Weak Forces}},
url = {http://arxiv.org/abs/2012.12260 https://link.aps.org/doi/10.1103/PhysRevLett.127.023601},
volume = {127},
year = {2021}
}

@article{Zhu2023,
author = {Shaocong Zhu and Zhenhai Fu and Xiaowen Gao and Cuihong Li and Zhiming Chen and Yingying Wang and Xingfan Chen and Huizhu Hu},
journal = {Photon. Res.},
number = {2},
pages = {279--289},
publisher = {Optica Publishing Group},
title = {Nanoscale electric field sensing using a levitated nano-resonator with net charge},
volume = {11},
month = {Feb},
year = {2023},
url = {https://opg.optica.org/prj/abstract.cfm?URI=prj-11-2-279},
doi = {10.1364/PRJ.475793},
}

\end{document}

\title{Supplementary material: Free expansion of a charged nanoparticle via electrostatic compensation}
\author{David Steiner}
\affiliation{Faculty of Physics, Vienna Center for Quantum Science and Technology (VCQ), University of Vienna, Boltzmanngasse 5, A-1090 Vienna, Austria}
\affiliation{Institute for Quantum Optics and Quantum Information (IQOQI) Vienna, Austrian Academy of Sciences, A-1090 Vienna, Austria}
\author{Yaakov Y. Fein}
\affiliation{Faculty of Physics, Vienna Center for Quantum Science and Technology (VCQ), University of Vienna, Boltzmanngasse 5, A-1090 Vienna, Austria}
\author{Gregor Meier}
\affiliation{Faculty of Physics, Vienna Center for Quantum Science and Technology (VCQ), University of Vienna, Boltzmanngasse 5, A-1090 Vienna, Austria}
\author{Stefan Lindner} 
\affiliation{Faculty of Physics, Vienna Center for Quantum Science and Technology (VCQ), University of Vienna, Boltzmanngasse 5, A-1090 Vienna, Austria}
\author{Paul Juschitz} 
\affiliation{Faculty of Physics, Vienna Center for Quantum Science and Technology (VCQ), University of Vienna, Boltzmanngasse 5, A-1090 Vienna, Austria}
\author{Mario A. Ciampini} 
\affiliation{Faculty of Physics, Vienna Center for Quantum Science and Technology (VCQ), University of Vienna, Boltzmanngasse 5, A-1090 Vienna, Austria}
\author{Markus Aspelmeyer}
\affiliation{Faculty of Physics, Vienna Center for Quantum Science and Technology (VCQ), University of Vienna, Boltzmanngasse 5, A-1090 Vienna, Austria}
\affiliation{Institute for Quantum Optics and Quantum Information (IQOQI) Vienna, Austrian Academy of Sciences, A-1090 Vienna, Austria}
\author{Nikolai Kiesel}
\affiliation{Faculty of Physics, Vienna Center for Quantum Science and Technology (VCQ), University of Vienna, Boltzmanngasse 5, A-1090 Vienna, Austria}

\let\clearpage\relax\maketitle

\section{Experimental details}
\begin{figure}[h]
    \centering
    \includegraphics[width=0.7\linewidth]{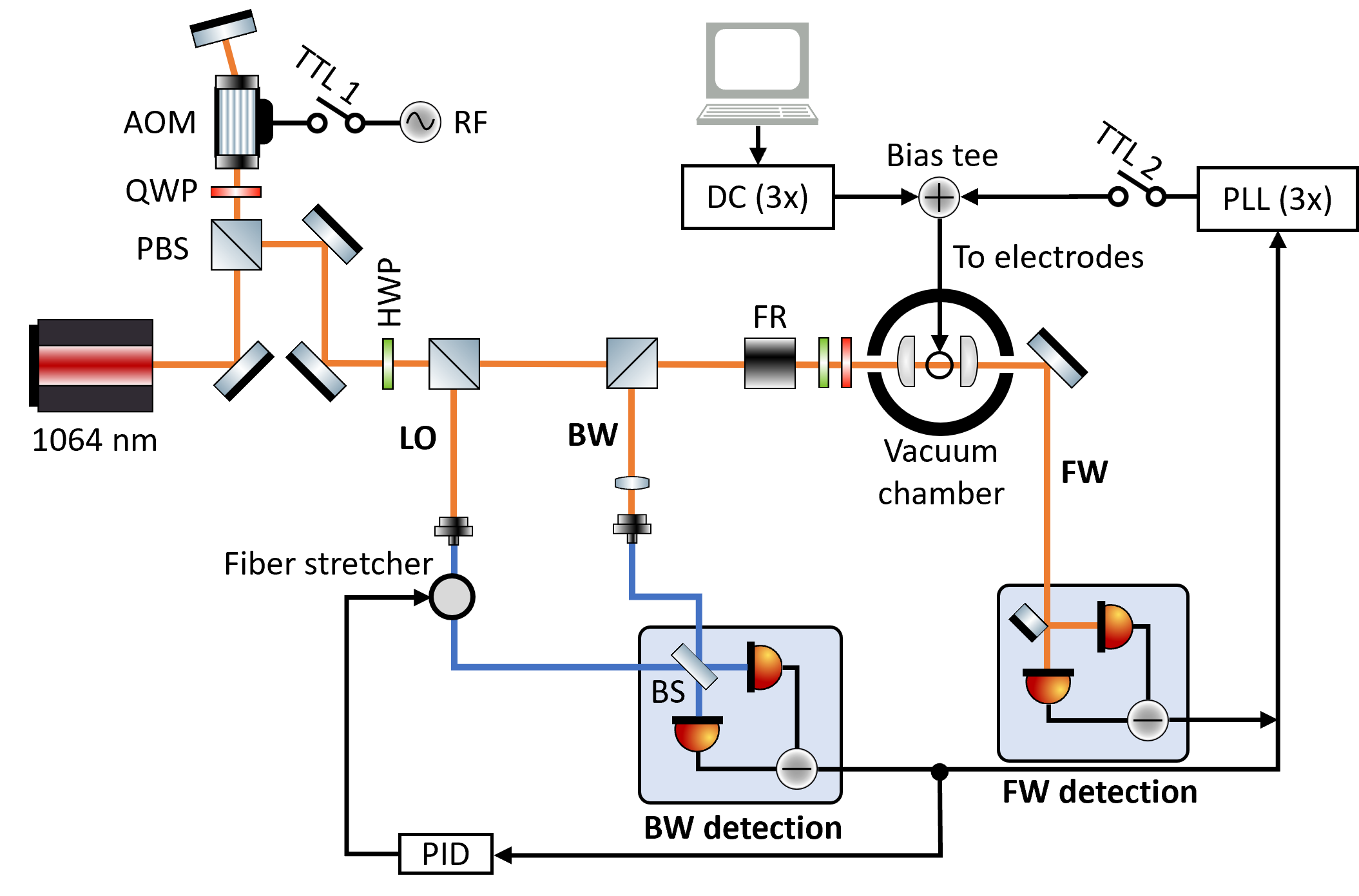}
    \caption{A schematic of the experimental setup, showing the trapping beam passing the double-pass AOM, the homodyne detection via mixing of a phase-locked local oscillator (LO) with the back-scattered light (BW), and one of the forward (FW) split detections. Orange lines indicate free-space optical paths, blue lines indicate in-fiber paths and black lines indicate electronic signals. PBS: Polarizing beamsplitter; BS: 50:50 non-polarizing beamsplitter; AOM: Acousto-optic modulator; QWP: quarter-wave plate; HWP: half-wave plate.}
    \label{fig:schematic}
\end{figure}

\vspace{-4mm}
A schematic of the experiment is shown in Fig.~\ref{fig:schematic}.
Switching of the feedback signals is done via three Mini-Circuits ZASWA-2-50DRA+ controlled with TTL pulses from a Keysight 33500B arbitrary waveform generator.
The trap is switched by passing the RF drive (Wieserlabs WL-FlexDDS-NG-DUAL) through a ZASWA switch before amplifying it (Mini-Circuits ZHL-03-5WF+) and sending it to the double-pass AOM.
The double-pass configuration ensures high extinction of the trapping beam during the pulsing.
The optical switching has rise/fall times of about 170\,ns with a 380\,ns delay (to the 90\%level) with respect to the TTL trigger, while the feedback is disabled within 50\,ns of the TTL. 

To mix the DC compensation voltages with the RF feedback signals, we use three identical bias tees each with a 0.94\,µF capacitor and a 2.2\,mH inductor.
Adding a 150\,k$\Omega$ resistor in series with the inductor avoids ringing of the DC bias after the feedback switching. 

The trap polarization is linear for the measurements, but we found that temporarily setting a polarization state closer to circular improved stability when pumping to UHV. 
In particular, rotational/librational modes are driven to higher frequencies, far from the translational modes of interest.

\section{Electrode configuration}
\begin{figure}[h]
    \centering
    \includegraphics[width=0.4\linewidth]{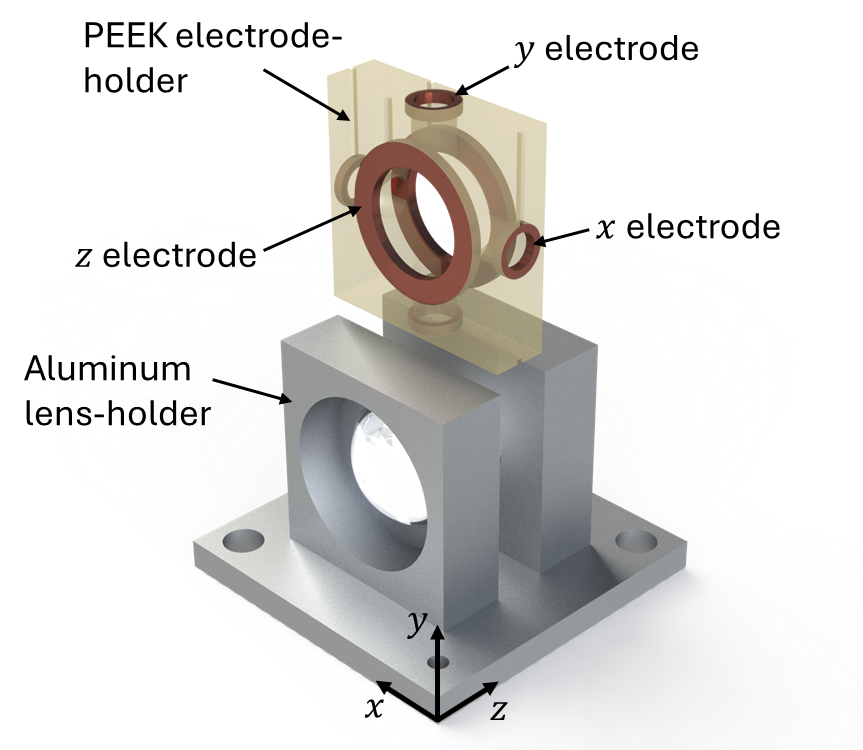}
    \caption{An exploded-view rendering of the electrode assembly and the lens-holder, with the electrode-holder semi-transparent for clarity. All electrodes are isolated and connected via a sub-D feedthrough; one electrode of each pair is labeled in the figure.}
    \label{fig:electrodeCAD}
\end{figure}

\vspace{-4mm}
The electrode configuration consists of six copper electrode rings press-fitted into an insulating PEEK holder, which is likewise press-fitted between the lenses, as shown in Fig.~\ref{fig:electrodeCAD}. 
This mounting enables reasonable orthogonality of the electrode pairs and symmetry around the trapping region.
The use of rings was for optical access of the trapping beam in $z$ and camera access in the radial directions.
The axial electrodes have an outer/inner diameter of 15\,mm/10.5\,mm and are 3.5\,mm apart, while the radial electrodes have an outer/inner diameter of 5\,mm/3.5\,mm and are spaced by approximately 20\,mm.

\section{Cross-cooling}
\begin{figure}[h]
    \centering
     \includegraphics[width=0.4\linewidth]{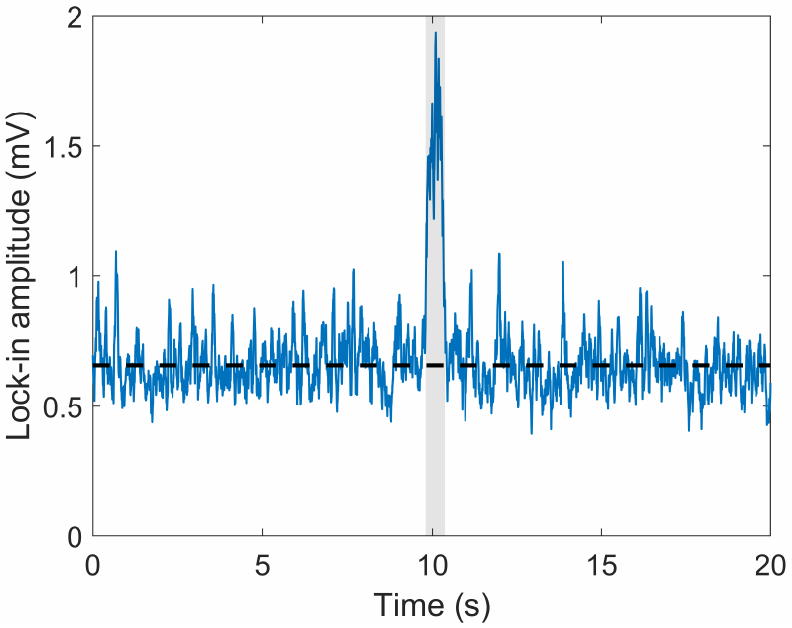}
    \caption{Demodulation of the $x$ detection with the phase-locked loop at $7\cdot10^{-2}$\,mbar. Until 10\,seconds, the $x$ motion is cooled via the $x$ electrodes, at which point the cooling is briefly disabled (gray region) and then re-enabled via the $y$ electrodes. The feedback amplitude is increased by a factor of about 4 compared to $x$ cooling to achieve a comparable cooling performance; these values enter the calculation of $C^{-1}$ as described in the text.}
     \label{fig:cross_cooling}
\end{figure}

In Fig.~\ref{fig:cross_cooling} we show a sample lock-in trace of the $x$ mode during the "cross-cooling" procedure used for determining the inverse of the projection matrix $C$.
In this example, we lock a phase-locked loop to the $x$ mode using the $x$ detection. 
We then apply this signal, appropriately phase-shifted, to the $x$ electrodes to perform moderate linear cooling of the $x$ mode (first half of the figure).
We then apply this feedback signal to the $y$ electrode instead and adjust the amplitude (and phase by 180°, if necessary) until the $x$ mode is cooled to the level achieved with the $x$ electrodes (second half of the figure).
In this condition, we have $C_{x'x}V_{x'}^x = C_{y'x}V_{y'}^x$, which, when we perform an analogous procedure for the remaining permutations of feedback/detection, allows us to rewrite $C^{-1}$ as

\vspace{-4mm}

\begin{equation}
    C^{-1} =
    \begin{pmatrix} 1 & {V_{x'}^x}/{V_{y'}^x} & {V_{x'}^x}/{V_{z'}^x} \\ {V_{y'}^y}/{V_{x'}^y} & 1 & {V_{y'}^y}/{V_{z'}^y} \\ {V_{z'}^z}/{V_{x'}^z} & {V_{z'}^z}/{V_{y'}^z} & 1 \end{pmatrix}^{-1}
    \begin{pmatrix} {1}/{C_{x'x}} & 0 & 0 \\ 0 & {1}/{C_{y'y}} & 0 \\ 0 & 0 & {1}/{C_{z'z}} \end{pmatrix}
    \coloneq \tilde{C}^{-1}S.
\end{equation}

The left matrix, $\tilde{C}^{-1}$, contains the required ratios of voltages for independent forces along the trap axes in its columns.
The matrix $S$ contains the diagonal elements of $C$ which are only required if calibrated forces are desired, and is straightforwardly obtained via harmonic drive measurements in a calibrated detection.
If only independent forces are required, then the column scaling is arbitrary and the method is completely calibration-free. 
In our setup, we find
\begin{equation}
    \tilde{C}^{-1} = 
    \begin{pmatrix}
    1.1 & 0.35 & -35 \\ 0.39 & 1.1 & 4.1 \\ 0.0011 & -0.0013 & 0.93
    %1 & 0.3158 & -37.2709 \\ 0.3634 & 1 & 4.3893 \\ 0.0011 & -0.0012 & 1
    \end{pmatrix}.
\end{equation}
Repeated measurements of the matrix entries showed minimal change over time. 

\section{Scaling with $\tau$}
\begin{figure}[h]
\centering
\begin{minipage}{.49\textwidth}
  \centering
  \includegraphics[width=0.8\linewidth]{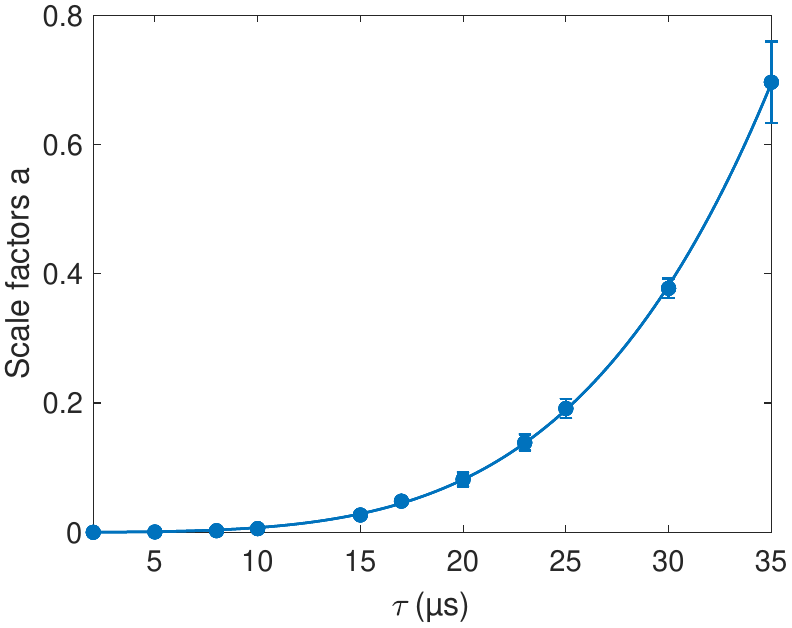}
\end{minipage}
\begin{minipage}{.49\textwidth}
  \centering
  \includegraphics[width=0.8\linewidth]{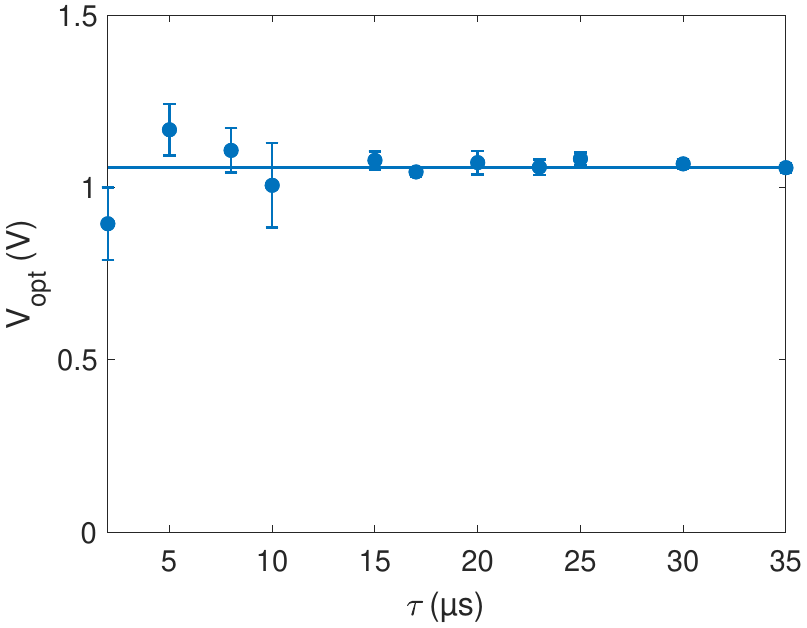}
\end{minipage}
\caption{
Compensation scans along $z$ as a function of $\tau$.  \textbf{Left:} Fitted parabola scale factors $a$ of each scan. The quartic dependence dominates at large $\tau$, as seen by the good agreement of a $\tau^2+\tau^4$ fit (blue line). Error bars are 95\% confidence intervals of the fitted parabola scale factors $a$.
\textbf{Right:} Fitted parabola minima $V_{\text{opt}}$ for each scan. The blue line is the mean of the $V_{\text{opt}}$ values to illustrate the lack of any $\tau$ dependence, as expected for compensation of a constant force. Error bars are 95\% confidence intervals of the fitted minima.}
\label{fig:t4_and_tau}
\end{figure}

\vspace{-4mm}
As described in the main text, the mean oscillation energy scales as $F_z^2\tau^4$ for $\tau\gg\Omega^{-1}$ in the presence of a constant force $F_z$. 
To see this dependence, we perform a series of 1D axial compensation scans with varying free evolution time $\tau$ and fit a parabola $a(V-V_{\text{opt}})^2+b$ to each curve.
Since the voltage is proportional to the force, we can see the expected $\tau^4$ dependence in the parabola scale factors $a$, as shown in Fig.~\ref{fig:t4_and_tau} (left).

We additionally plot the behavior of the fitted minima $V_{\text{opt}}$ as a function of $\tau$ in Fig.~\ref{fig:t4_and_tau} (right).
The lack of any non-constant trend implies that the force being compensated is nearly constant, for which one expects a $\tau$-independent compensation voltage.
Likewise, it confirms that there is no significant fixed momentum kick from switching of the trap/feedback, which would be indicated by an asymptotic convergence of such a plot. 

\section{Charge characterization and control}
\vspace{-4mm}
Particle charge control is performed via a high-voltage electrode discharge in the vicinity of the particle at medium vacuum ($\approx$1\,kV, mid $10^{-1}$\,mbar typical)~\cite{Frimmer2017}. 
The charge is measured by applying an electronic sinusoidal drive at 120\,kHz and demodulating the response in the homodyne detection.
The absolute charge state is then calibrated by taking a histogram of single-e charge step jumps in the demodulated detection.

We determine the force transduction coefficients $C_{NV} = F/V$ for each axis by measuring the response of a harmonic drive as a function of the drive amplitude in a calibrated detection.
The detector calibration is performed via equipartition at high pressure, as is now standard for levitated systems~\cite{Magrini2021,Hebestreit2018_calib}.
We find $C_{NV}$ values of order $10^{-16}$\,N/V for the axial electrodes and $10^{-18}$\,N/V for both radial electrodes, with the difference due to the geometry of the electrode arrangement.

Minimizing the environmental charge state is desirable, both to reduce the required compensation voltages and potentially reduce the effect of a fluctuating force component.
With environmental charge control we aim to minimize the background electrostatic forces due to charged dielectric surfaces while keeping the particle charge state unchanged..
We investigated using a flow of nitrogen ionized via a glowing tungsten filament mounted directly in the chamber as tested for the discharge of LIGO test masses~\cite{Ugolini2011}, but it showed only a moderate effect.
A general problem we encountered was that any effect of charging/discharging at high pressures is then obscured by the random change in charge state typically observed on pumping, which we attribute to triboelectric charging effects.

\begin{figure}[h]
    \centering
    \includegraphics[width=0.4\linewidth]{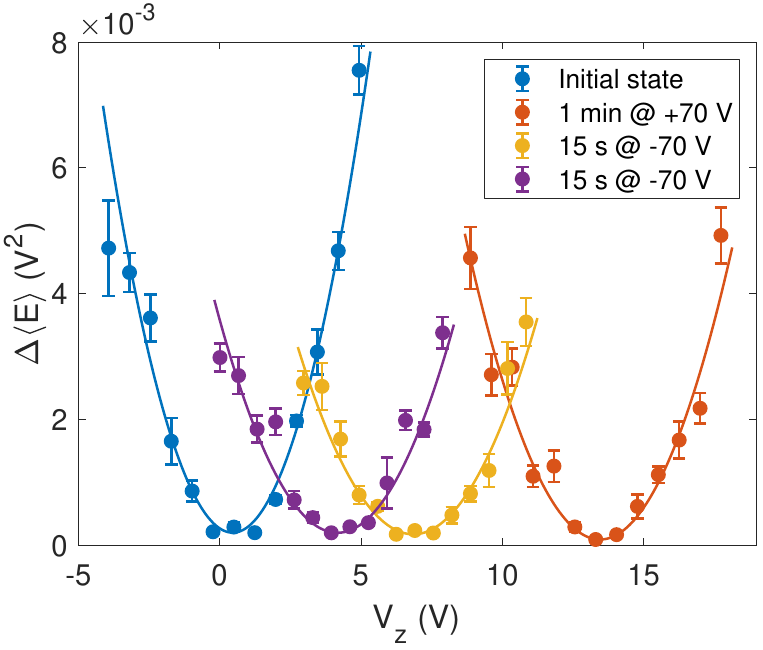}
    \caption{Environmental charge control via a glowing filament and electron steering. The environmental charge state along $z$ is measured via a series of 10\,µs compensation scans. $V_{\text{opt}}$ was initially close to 0\,V (blue) before the filament was "flashed" at 1.5\,A for one minute with the $z$ electrodes held at 70\,V (other electrodes grounded), which brought $V_{\text{opt}}$ to 14\,V (red). Subsequent 15 second flashes with the $z$ electrode held at -70\,V brought $V_{\text{opt}}$ to 7\,V (yellow) and then 4\,V (purple).    
    }
    \label{fig:lightbulb}
\end{figure}

To avoid the random environmental charging associated with pumping, we developed a vacuum-compatible method.
By running 1.5\,A through the tungsten filament at low pressures ($10^{-7}$\,mbar or below), we produced a flux of low energy electrons, which we could then steer to various surfaces by biasing the electrodes.
This procedure is demonstrated in Fig.~\ref{fig:lightbulb}, where we use a series of compensation scans to show how we manipulate the background electrostatic force along $z$.

\begin{figure}[h]
    \centering
    \includegraphics[width=0.4\linewidth]{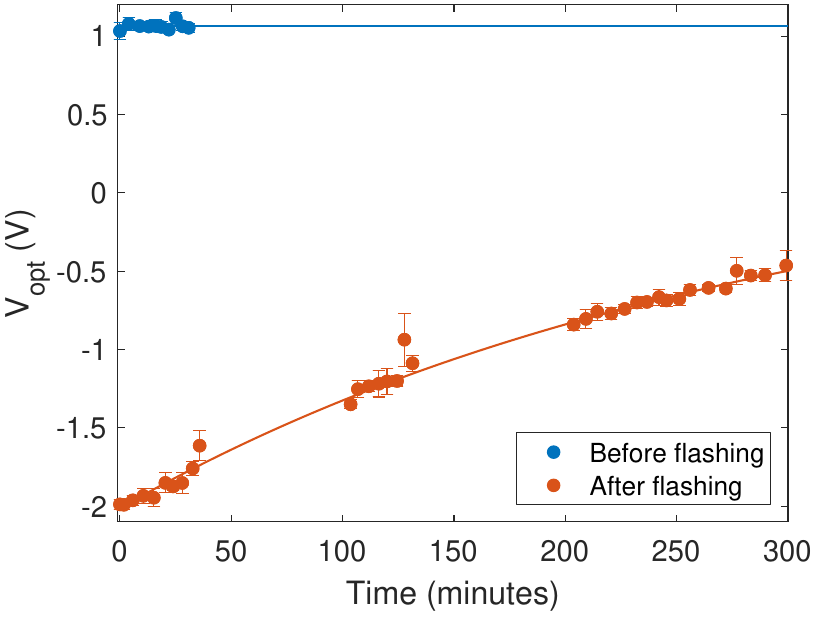}
    \caption{Stability of the optimal compensation voltage over time. In blue, we plot the $V_{\text{opt}}$ extracted from a series of $z$ compensation scans repeated over about 40 minutes ($\tau=20$\,µs) without "flashing" the filament, with the blue line the mean value. The red points were taken after a single flash, showing a drift of $V_{\text{opt}}$ which we fit to $V_f+V_0e^{-t/RC}$ to find $V_f\approx0.3$\,V and $RC\approx 300$\,min (red line). The pressure dropped from $1\cdot10^{-7}$\,mbar to $7\cdot10^{-8}$\,mbar over the course of the measurements. Error bars are 95\% confidence intervals of the fitted minima.}
    \label{fig:min_vs_time}
\end{figure}

It is also critical that the resulting environmental charge state remains stable over the timescale of any subsequent measurements. 
In Fig.~\ref{fig:min_vs_time}, we plot the measured values of $V_{\text{opt}}$, showing that the mean value of the force remains constant over hour timescales (blue points). 
The standard deviation of 0.02\,V corresponds to a force variation of order $10^{-18}$\,N.
However, we observed that the flashing procedure creates a less stable charge environment (red points), as shown in Fig.~\ref{fig:min_vs_time}, presumably as the deposited charges eventually find a route to ground through the dielectric electrode holder.

\section{Reheating and recompression}
\begin{figure}[h]
    \centering
    \includegraphics[width=0.4\linewidth]{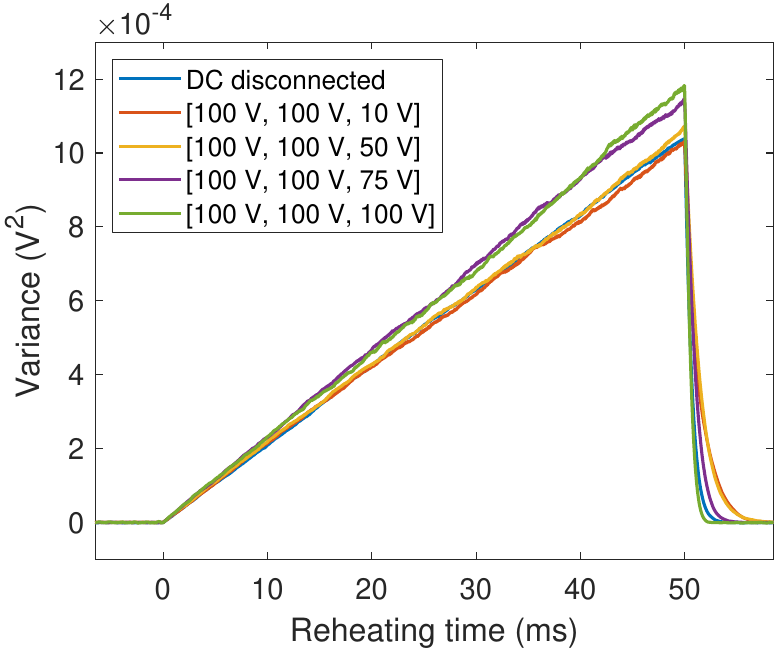}
    \caption{Reheating measurements for various bias voltages $[V_x,V_y,V_z]$. An increase above the reference measurement with no bias is only observed for $z$ electrode voltages above 50\,V, implying that voltage noise of our electronics does not yet play a relevant role.}
    \label{fig:reheating}
\end{figure}

\vspace{-4mm}
Reheating measurements at various DC bias levels are shown in Fig.~\ref{fig:reheating}.
The measurement consists of disabling the $z$ feedback for 50\,ms and measuring the growth in the variance of the timetrace using a moving variance with a window of 40\,µs.
Each curve is the average of 2250 repetitions (4500 repetitions for the reference measurement with DC bias disconnected) at pressures below 6$\cdot10^{-8}$\,mbar.
The measurements indicate that voltage noise only becomes significant (as compared to the near-recoil limited rate at these pressures) when the axial voltages are held above 50\,V, over an order of magnitude beyond typical compensation voltages.
Furthermore, the measurements with 10\,V axial bias were performed using the 100\,V output range of the SRS DC205 supply, for which the RMS noise is significantly higher than in the 10\,V or 1\,V range.
This implies that the increased heating rate observed at 75 and 100\,V is not due to the range-related RMS noise, since this depends only on the range setting and not the absolute voltage, which we independently verified.
The observation demonstrates the need for measuring non-optical noise sources in such experiments, especially in Paul or hybrid Paul-optical trap platforms where large DC voltages are often employed. 

\begin{figure}[h]
    \centering
    \includegraphics[width=0.4\linewidth]{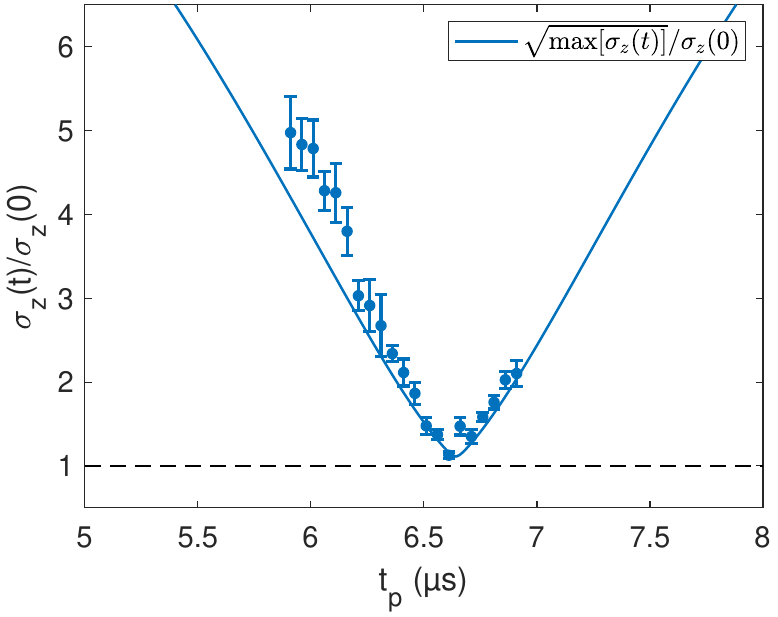}
    \caption{Recompression protocol with $\tau=5$\,µs and 150 repetitions per point. Recoil and gas diffusion are included in the theory (blue line), which brings the minimum $\sigma_z$ to slightly above the original value. We subtract 90\,ns from the experimental $t_p$ values for a better agreement with the theory, which we attribute to a small difference in the AOM rise and fall times.}
    \label{fig:recomp}
\end{figure}

As an additional coherence check, we perform a recompression protocol.
This consists of a free evolution for a time $\tau$, followed by a pulse for a time $t_p$, followed by another free evolution of duration $\tau$.
One should have recompression of the position variance for $t_p = 2\theta/\Omega+n\pi/\Omega_z$ for $n\in\mathbb{N}$, with $\theta$ the phase space ellipse's rotation angle to the position axis after the time $\tau$.
In other words, $\sigma^2_z(2\theta/\Omega+n\pi/\Omega) = \sigma^2_z(0)$ for a perfectly coherent protocol.

In Fig.~\ref{fig:recomp}, we plot such a recompression protocol as a function of $t_p$ for a fixed $\tau=5$\,µs, showing the standard deviation return to nearly its initial value (here for $n=1$).
We use a similar analysis as for Fig. 5 of the main text (see Analysis Details below), but use 100\,µs of data after the second free evolution and a two-stage fitting including all three modes for a more robust estimate of the axial motion.  

\section{Required compensation accuracy}
\vspace{-4mm}
\begin{figure}[h]
    \centering
    \includegraphics[width=0.4\linewidth]{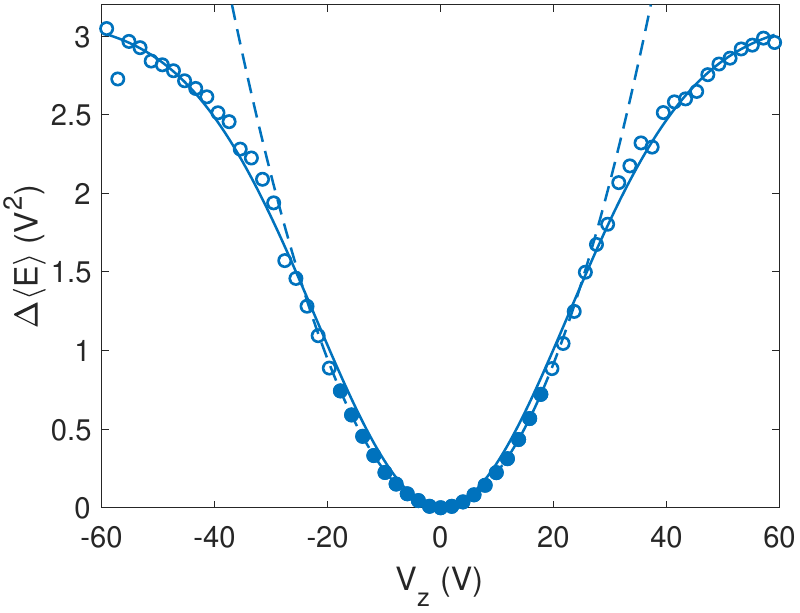}
    \caption{A compensation scan along $z$ with $\tau=15$\,µs and a much larger than typical voltage range to intentionally probe trap nonlinearities. The solid line is a gaussian fit and the  dashed line is a parabolic fit to the filled-in points, showing a significant onset of nonlinearities from about 30\,V.}
    \label{fig:big_z}
\end{figure}

A mean displacement of the particle wavepacket during free evolution that results in an uncontrolled probing of trap nonlinearities is undesirable.
We therefore experimentally map out the onset of these nonlinearities by performing a compensation scan along $z$ with a large voltage range, as shown in Fig.~\ref{fig:big_z}.
One can then convert the voltage axis to displacement via $\Delta z = C_{NV} (V-V_{\text{opt}})\tau^2/2m$, from which we determine that significant nonlinearities set in for displacements above approximately 170\,nm. 

Using the radial and axial trap waists determined from the trapping frequencies, we can relate the onset of nonlinearities axially to a corresponding radial displacement.
In our trap, equivalent nonlinearities will be probed for a 2.9 times smaller radial displacement than axial.
Converting voltage to displacement as above, we find that to avoid probing nonlinearities after a 100\,µs free evolution, we require about 1\,V accuracy axially and 20\,V accuracy radially for the compensation voltages.

\section{Analysis details}
\vspace{-4mm}
For the 1D compensation scans, we take the variance of the timetrace (no filtering beyond an analog 1\,MHz low-pass filter) over several oscillation periods (typically 40\,µs).
The variance over several periods is proportional to the amplitude of the oscillations squared and therefore the mean energy.
Since the back-scattered homodyne detection is orders of magnitude more sensitive to the $z$ motion than the radial, this is to a good approximation the energy of the $z$ mode; the other modes can be extracted via a power spectral density analysis as done in Fig. 2 of the main text.
We generally leave these curves uncalibrated (units of $V^2$) since they are only used for determining the minimum of the parabola, $V_{\text{opt}}$.

For Figures 4 and 5 of the main text, we instead fit 25\,µs of post-recapture data to $a\sin(\Omega_zt+\phi)+bt+c$ to extract the amplitude and phase of each oscillation in a calibrated detection (the linear drift accounts for a small drift in the homodyne DC level).
We find no significant difference between fitting just the axial mode vs. all three translational modes; in general the presented results are very robust to the details of the fitting procedure.
After fitting each trajectory, we take the mean of the squared amplitudes to obtain the mean energy, and the variance of the fitted trajectories to obtain the ensemble variance.
The position standard deviation plotted in Figure 5 of the main text corresponds to the maximum value of this ensemble variance, i.e., the point during the trap evolution when the phase space ellipse is aligned with the position axis (see e.g. Fig. 1 of the main text). 
Error bars for the position standard deviation are estimated by sub-dividing the 150 repetitions into 10 bins, taking the statistics of each subsample and then the standard error of the mean. 

\bibliography{references}